\documentclass[twocolumn,twoside,slac_two]{revtex4}
\usepackage{graphicx}
\usepackage{fancyhdr}
\usepackage{graphics}
\usepackage{epstopdf}
\usepackage{textpos}
\usepackage{slashed}
\newcommand{\mad}{\texttt{MadGraph }}
\newcommand{\pyt}{\texttt{PYTHIA }}

\begin{document}

\title{\centering W, Z + Jets production with CMS detector}
\author{
\centering
\begin{center}
Sandro Gonzi \\on behalf of the CMS Collaboration
\end{center}}
\affiliation{\centering Universit\`{a} degli Studi di Firenze and INFN Sezione di Firenze, via G. Sansone 1 - 50019 - Sesto Fiorentino (FI), Italy}
\begin{abstract}
We present a study of jet production in association with W and Z bosons in proton-proton collisions at a centre-of-mass energy of 
$7\ \mathrm{TeV}$ using the full 2010 data set collected by CMS corresponding to an integrated luminosity of $(35.9 \pm 1.4)\ \mathrm{pb^{-1}}$. We report the measurement of ratios $\sigma(\mathrm{V}\ +\ \geq\ n\ \mathrm{jets})/\sigma(\mathrm{V})$ and $\sigma(\mathrm{V}\ +\ \geq\ n\ \mathrm{jets})/\sigma(\mathrm{V}\ +\ \geq\ (n - 1)\ \mathrm{jets})$, where V represents either a W or a Z and $n$ stands for number of jets. 
\end{abstract}

\maketitle
\thispagestyle{fancy}


\section{INTRODUCTION}
The study of the production of vector bosons W and Z in association with hadronic jets (known as ``V + jets'' process), provides a
stringent test of perturbative QCD calculations. Moreover, this process is a significant source of background in searches for new physics, Standard Model Higgs boson and for studies of the top quark. A precise measurement of the cross section and an understanding of its kinematics is then essential. Finally, a clear signature associated at a relatively well known physics make this process an useful test for the commissioning of the detectors.

We present the results of the analysis obtained with the full 2010 data sample collected by the Compact Muon Solenoid (CMS) experiment at the Large
Hadron Collider (LHC) in proton-proton collisions at a centre-of-mass energy of $7\ \mathrm{TeV}$ and corresponding to an integrated luminosity of $(35.9 \pm 1.4)\ \mathrm{pb^{-1}}$. To reduce systematic uncertainties, we measure the V + $n$ jets cross sections relative to the inclusive W and Z cross sections. 
The complete V + jets analysis is reported in \cite{bib:PAS}.

\section{EVENT SELECTION}

\begin{figure*}[t]
\centering
\includegraphics[width=65mm]{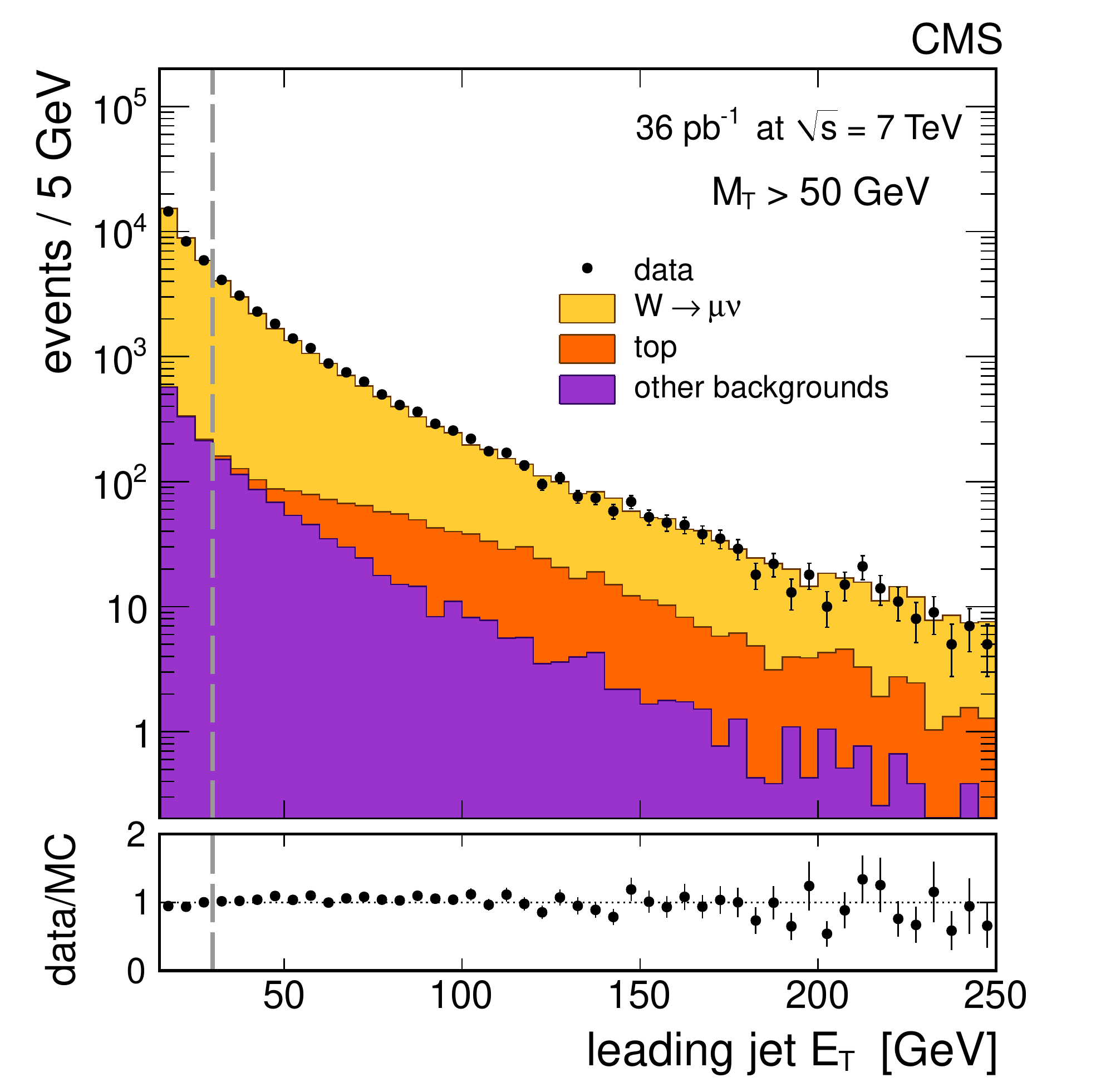} \qquad \qquad \qquad
\includegraphics[width=65mm]{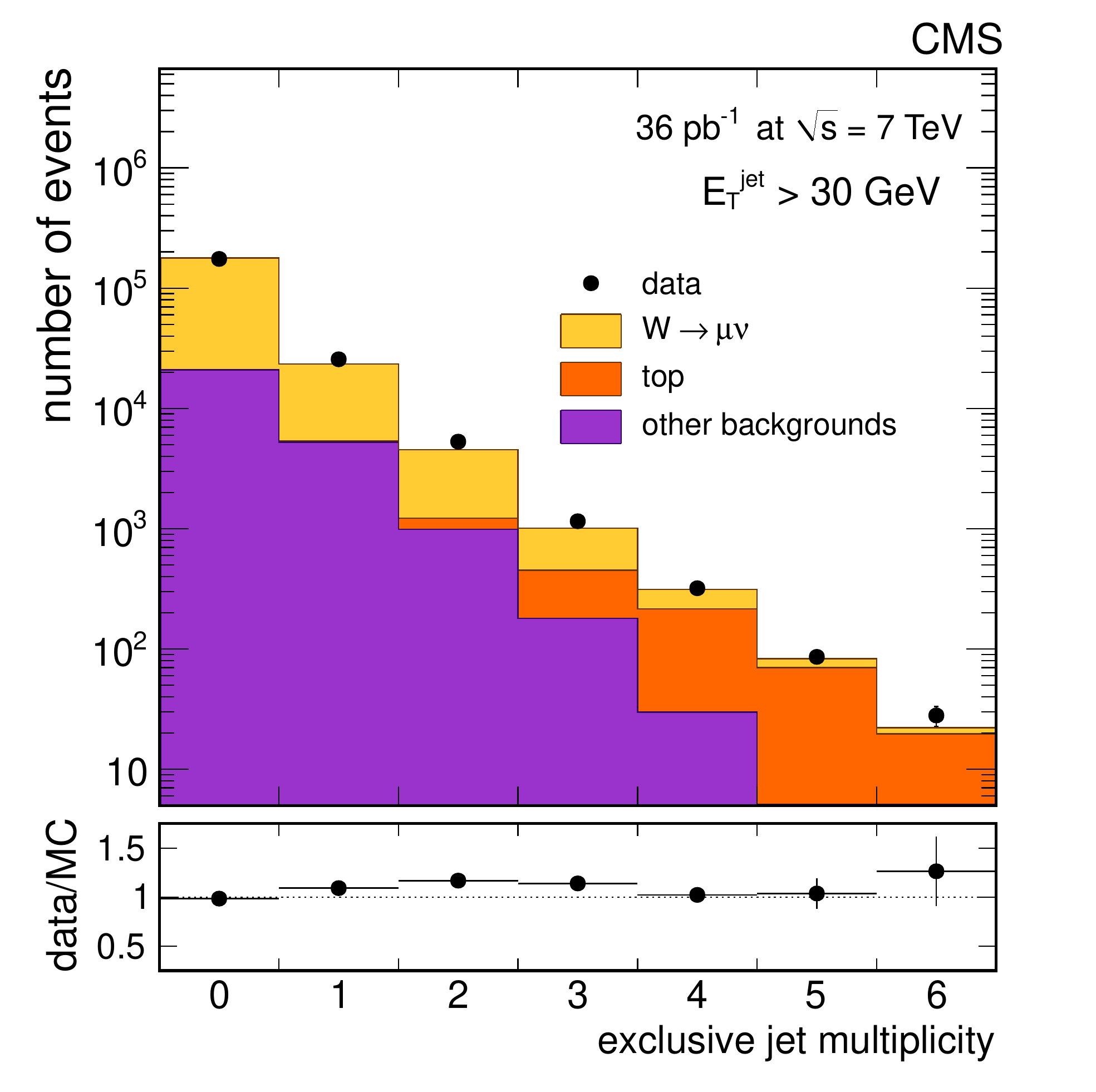}
\caption{Distributions of the $E_T$ for the leading jet (left) and of the number of reconstructed jets (right) in events $\mathrm{W} \rightarrow \mu \nu$. The ratio between the data and the simulation is also shown. The line at $p_T = 30\ \mathrm{GeV}$ corresponds to the threshold imposed for counting jets.}
\label{fig:1}
\end{figure*}

A hardware-based trigger system selects a sample of electrons and muons then filtered in the online cluster with algorithms that have evolved following the rapid rise of the LHC luminosity. The largest sample of leptons preselected are electrons with $p_T \geq 17\ \mathrm{GeV}$ and muons with $p_T \geq 15\ \mathrm{GeV}$.

Monte Carlo (MC) simulation samples with a W or a Z boson are generated with the \mad event generator, producing parton-level events with a vector boson and up to four jets on the basis of a matrix-element calculation. \mad is interfaced to the \pyt program for parton shower simulation, modified in such a way that the hardest emission is modeled using the exact matrix element calculation for one additional real emission.
Top background processes are generated with \mad while QCD multi-jet and $\gamma$ + jets processes are generated with \pyt alone.
Whenever available, a NNLO or NLO normalization is performed.

Minimum-bias events are superimposed to the hard interaction to simulate the event pile-up observed in data.

A Particle Flow (PF) algorithm is used to reconstruct both the missing transverse energy ($\slashed{E}_T$) and the jets in the event. The PF algorithm creates a complete event description by collecting information from all of the sub-detectors and linking it together. Objects are formed into the categories of muons, electrons, photons, charged hadrons and neutral hadrons. 

Signal selection begins with the identification of a so-called ``leading lepton'' following the standard established by the measurement of the inclusive W and Z cross sections \cite{bib:signal}.

For electron candidates we require a transverse momentum selection $p_T > 20\ \mathrm{GeV}$ and that the position of the ECAL supercluster lies in a fiducial region identified by $|\eta| < 2.5$ with $1.4442 < |\eta| < 1.566$ excluded (this allow to reject electrons close to the barrel/endcap region transition where cables and services compromise the reconstruction). The selected electron must match the object that triggered the event readout and it must also pass tight quality requirements (including identification, isolation, and conversion rejection) which correspond to a lepton efficiency of roughly 80\% as evaluated with a \mad + \pyt simulated sample.
If a second electron of $p_T > 10\ \mathrm{GeV}$, detected within the ECAL fiducial region, passes a looser set of quality cuts (corresponding to an efficiency of about 95\%) and its invariant mass with the leading electron is in the range $[60-120]\ \mathrm{GeV}$, then the event is assigned to the Z + jets sample. Otherwise, the event is assigned to the W + jets sample. 
Events with a muon with $p_T > 15\ \mathrm{GeV}$ and $|\eta | < 2.4$ are rejected from the W + jets sample to reduce $t\bar{t}$ contamination.

For muon candidates we require the presence of an isolated high quality muon with $p_T > 20\ \mathrm{GeV}$ in the region $|\eta| < 2.1$ with a transverse impact parameter $|d_{xy} | < 2\ \mathrm{mm}$ to suppress cosmic-ray muon background. The isolation requirement is obtained by requiring that the combined activity of the tracker and calorimeters around the muon relative to the muon $p_T$ is less than 0.15.
If a second muon of $p_T > 10\ \mathrm{GeV}$ is detected in the region $|\eta | < 2.5$ such that the di-muon invariant mass 
lies in the range $[60-120]\ \mathrm{GeV}$, then the event is assigned to the Z + jets sample. Otherwise, the event is assigned to the W + jets sample

For the W + jets samples, the transverse mass $M_T$ is computed starting from the lepton missing transverse energy $\slashed{E}_T$ using the relation \ $M_T = \sqrt{2 p_T \slashed{E}_T (1-\cos \Delta \phi)}$, where $\Delta \phi$ is the angle on the plane orthogonal to the direction of the beams between the direction of the $p_T$ lepton and the $\slashed{E}_T$ one. To avoid a region at low $M_T$ containing essentially no signal we require that $M_T > 20\ \mathrm{GeV}$.

Jets are reconstructed from the particle collection created with the PF algorithm and are formed with the anti-$k_T$ clustering algorithm \cite{bib:anti_kt} with a size parameter of $R = 0.5$. Jet energy corrections (JEC) are applied to flatten the jet energy response as a function of $\eta$ and $p_T$.
Jets must first satisfy identification criteria to eliminate those originating from noise in the calorimeter.
We requested $|\eta | < 2.4$ so that the jets fall within the tracker acceptance and, for the W sample, a $M_T > 50\ \mathrm{GeV}$ selection to reduce backgrounds.

Selected events are assigned to exclusive bins of jet multiplicity by counting the number of jets with $p_T > 30\ \mathrm{GeV}$.
The observed distributions for the leading jet transverse momentum and for the exclusive number of reconstructed jets in the W 
samples are shown in Figure \ref{fig:1} in data and simulation. 
The data is in good agreement with the \mad predictions normalized to the NNLO cross sections. 

\section{SIGNAL EXTRACTION}
\begin{figure*}[t]
\centering
\includegraphics[width=65mm]{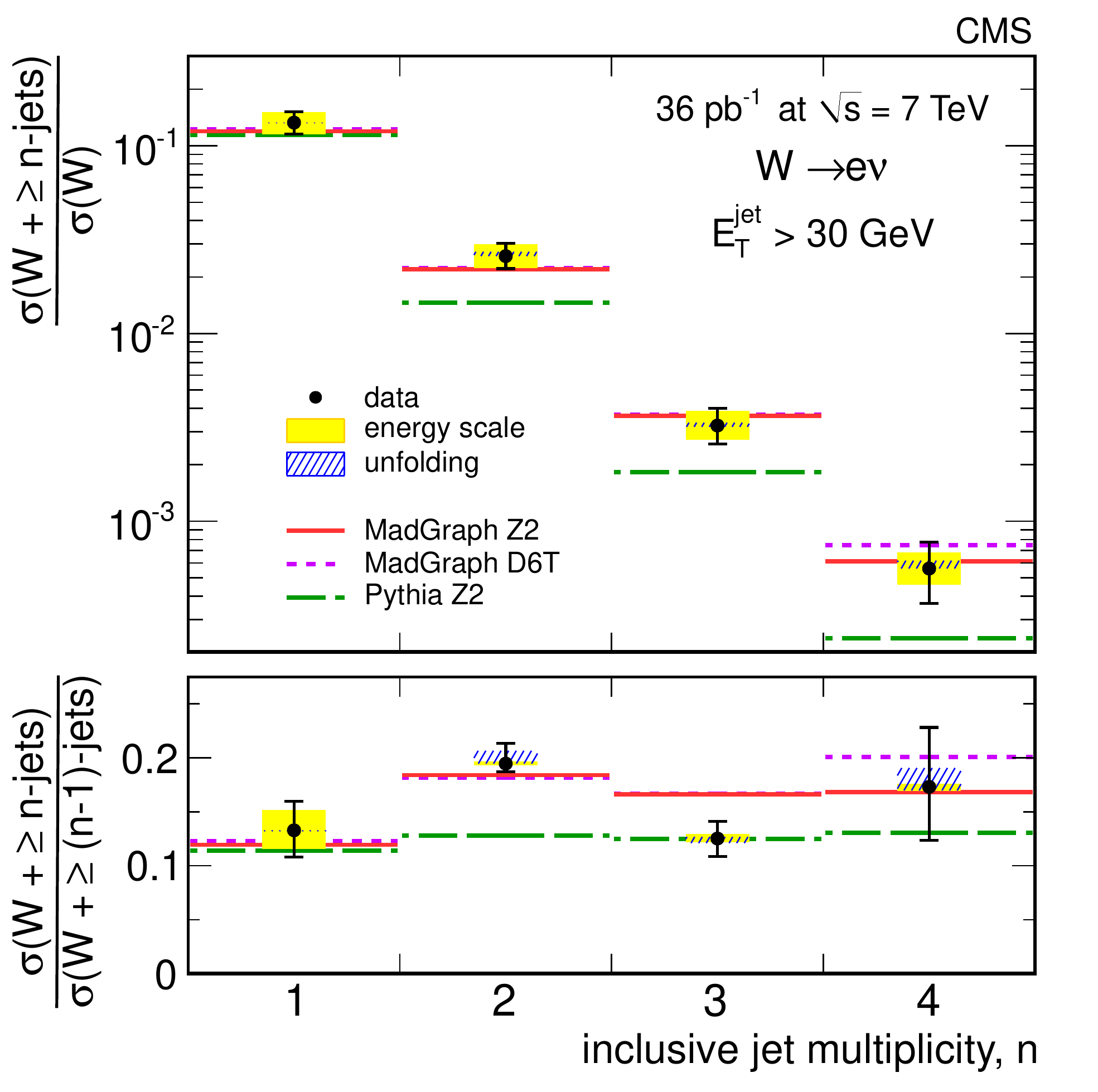} \qquad \qquad \qquad
\includegraphics[width=65mm]{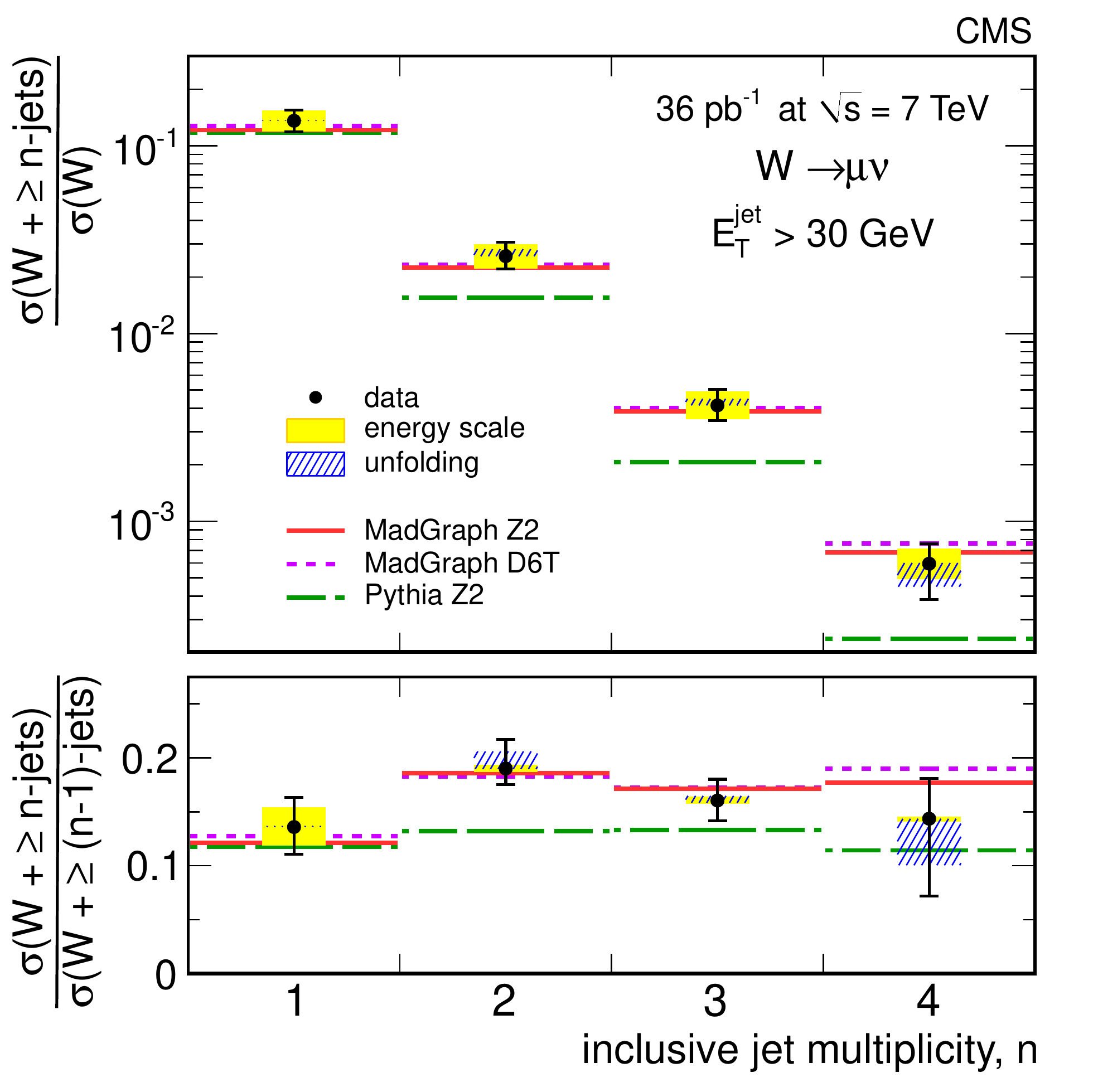}
\caption {The ratio $\sigma(\mathrm{W}\ +\ \geq\ n\ \mathrm{jets})/\sigma(\mathrm{W})$ and $\sigma(\mathrm{W}\ +\ \geq\ n\ \mathrm{jets})/\sigma(\mathrm{W}\ +\ \geq\ (n - 1)\ \mathrm{jets})$ in the electron channel (left) and muon channel (right) compared to expectations from \mad (Z2 and D6T tune) and \pyt (Z2 tune).}
\label{fig:2}
\end{figure*}
In order to provide model-independent results, we quote all results within the lepton and jet acceptance, and only correct for efficiency of the selection. 
The lepton efficiencies calculation is obtained from Z/$\gamma$* + jets data samples by means of a tag-and-probe method, starting from selected events with two lepton candidates of invariant mass in the range $[60-120]\ \mathrm{GeV}$. One lepton candidate, called the ``tag'', satisfies all selection requirements. The other one, called the ``probe'', is selected with criteria that depend on the efficiency being measured.
The signal yields are for two exclusive subsamples of events in which the probe lepton passes or in which it fails the selection criteria considered.
Fits are performed to the invariant-mass distributions of the pass and fail subsamples to extract the Z signal events. The measured efficiency is calculated from the relative level of signal in the pass and fail subsamples.
The lepton selection efficiency is the product of the reconstruction efficiency, the identification and isolation efficiency, and the trigger efficiency, each of which is calculated as a function of the jet multiplicity in the event.

For electrons, the efficiencies are roughly 70\% (60\%) for the W + jets (Z/$\gamma$* + jets) signal events.

For muons, the average efficiency is close to 85\% (86\%) for the W + jets (Z/$\gamma$* + jets) signal events and it exhibits a dependence on the jet multiplicity.

The signal yield is estimated using an extended likelihood fit to the di-lepton invariant mass $M_{l^{+}l^{-}}$ for the Z + jets sample and to $M_T$ for the W + jets sample. 

For the Z event samples, the main background processes, dominated by $t \bar{t}$ and W + jets, have little influence and do not produce a peak in the $M_{l^{+}l^{-}}$ distribution. In the $M_{l^{+}l^{-}}$ distribution we can then distinguish two components, one for the signal and one for all background processes.

For the W sample, background contributions can be divided into two components, one which exhibits a peaking structure in $M_T$, dominated by $t \bar{t}$, and another which does not, dominated by QCD multi-jet events. We perform a two-dimensional fit to the $M_T$ distribution and the number of $b$-jets, $n_{\mathrm{jet}}^{b\mathrm{-tagged}}$. The $M_T$ distribution allows the statistical separation of the signal from the non-peaking backgrounds, while $n_{\mathrm{jet}}^{b\mathrm{-tagged}}$ distinguishes the signal and the other backgrounds from $t \bar{t}$. 

In order to estimate the scaling rule of jets at the particle jet level, we apply an unfolding procedure that removes the effects of jet energy resolution and reconstruction efficiency. 

The main sources of systematic uncertainties in jet counting are the determination of the jet energy and jet energy resolution. We apply the jet energy scale (JES) corrections, available as a function of $\eta$ and $p_T$, to account for the detector response and inhomogeneities.
We apply also a pile-up energy correction to revise the pile-up subtraction method used that systematically removes $500\ \mathrm{MeV}$ to jets in events without pile-up. While the systematic uncertainty in the jet counting is correlated among the different jet multiplicities, the uncertainties from efficiency and fits are not. All statistical and systematic uncertainties are propagated in the unfolding procedure also estimating the systematic error associated with it.

\section{RESULTS}
From the unfolded exclusive jet multiplicity distributions we derive inclusive jet multiplicities and calculate two sets of ratios: 
$\sigma(\mathrm{V}\ +\ \geq\ n\ \mathrm{jets})/\sigma(\mathrm{V})$ and $\sigma(\mathrm{V}\ +\ \geq\ n\ \mathrm{jets})/\sigma(\mathrm{V}\ +\ \geq\ (n - 1)\ \mathrm{jets})$, where $\sigma(V)$ is the inclusive cross section. The results for the W sample are reported in Figure \ref{fig:2}, where the systematic uncertainties associated with the JES and the unfolding are shown as error bands. For a large number of jets, the \pyt simulation alone fails to describe the data, while the \mad + \pyt simulation agrees well, as expected.

\section{SUMMARY}
We measured the rate of jet production in association with a W or Z vector boson using the full 2010 dataset collected by CMS in \textit{pp} collision data at $\sqrt{s} = 7\ \mathrm{TeV}$. We also measured the ratios of cross sections $\sigma(\mathrm{V}\ +\ \geq\ n\ \mathrm{jets})/\sigma(\mathrm{V})$ and $\sigma(\mathrm{V}\ +\ \geq\ n\ \mathrm{jets})/\sigma(\mathrm{V}\ +\ \geq\ (n - 1)\ \mathrm{jets})$ where $n$ is the number of jets. All results agree well with simulations based on \mad + \pyt. 

\bigskip 

\end{document}